\documentclass{iopart}
\usepackage{graphicx}
\usepackage{iopams}

\begin{document}
\title[Pion HBT radii from RHIC to LHC]{Evolution of pion HBT radii from 
       RHIC to LHC -- Predictions from ideal hydrodynamics}
\author{Evan Frodermann, Rupa Chatterjee and Ulrich Heinz}
\address{Department of Physics, The Ohio State University, Columbus, 
         OH  43210, USA}

\begin{abstract}
We present hydrodynamic predictions for the charged pion HBT radii for a 
range of initial conditions covering those presumably reached in Pb+Pb 
collisions at the LHC. We study central ($b{\,=\,}0$) and semi-central 
($b{\,=\,}7$\,fm) collisions and show the expected increase of the HBT radii 
and their azimuthal oscillations. The predicted trends in the oscillation 
amplitudes reflect a change of the final source shape from out-of-plane to 
in-plane deformation as the initial entropy density is increased.\\[-1cm]
\end{abstract}

%%%%%%%%%%%%%%%%%%%%%%%%%%%%%%%%%%%%%%%%%%%%%%%%%%%%%%%%%%%%%%%%%%%%%%%%
\section{Introduction}\vspace*{-1mm}
%%%%%%%%%%%%%%%%%%%%%%%%%%%%%%%%%%%%%%%%%%%%%%%%%%%%%%%%%%%%%%%%%%%%%%%%
Ideal fluid dynamics has been used successfully to reproduce many aspects 
of heavy-ion collisions at RHIC \cite{Kolb:2003dz}. However, even though 
the hydrodynamic model correctly describes the hadron spectra at low 
transverse momenta, it fails to reproduce the transverse HBT radii 
($R_s$, $R_o$) at RHIC \cite{Heinz:2002un,Lisa:2005dd,Frodermann:2006sp}. 
It does, however, yield the correct normalized oscillation amplitudes for 
$R_s$ and $R_o$ \cite{Retiere:2003kf,Adams:2003zg,concepts}. We use it here 
to predict the expected trends for the evolution of the HBT radii at 
mid-rapidity in $(A{\approx}200)+(A{\approx}200)$ collisions from RHIC to 
LHC, including their normalized oscillation amplitudes in non-central 
collisions. These trends may be trustworthy, in spite of the model's 
failure to correctly predict the HBT radii at RHIC.

%%%%%%%%%%%%%%%%%%%%%%%%%%%%%%%%%%%%%%%%%%%%%%%%%%%%%%%%%%%%%%%%%%%%%%%%
\section{RHIC $\rightarrow$ LHC}\vspace*{-1mm}
%%%%%%%%%%%%%%%%%%%%%%%%%%%%%%%%%%%%%%%%%%%%%%%%%%%%%%%%%%%%%%%%%%%%%%%%
Extending the hydrodynamic code AZHYDRO \cite{Kolb:2000sd} 
from RHIC conditions to the LHC regime requires a recalculation of the 
Glauber initial conditions to reflect the achieved higher initial energy 
density and temperature. Some of the expected trends were outlined in 
\cite{Heinz:2002sq} where the initial temperature was increased to 2\,GeV 
to show the possibility of a transition from out-of-plane to in-plane 
deformation of the freeze-out source in non-central Pb+Pb collisions.
Such extreme initial temperatures are, however, not likely to be reached 
in Pb+Pb collisions at the LHC. We here present predictions for a more 
realistic range of initial conditions and a better equation of state (EoS) 
than the ideal guark-gluon gas without phase transition used in 
\cite{Heinz:2002sq}.

Hydrodynamics cannot predict its own initial conditions as a function of 
$\sqrt{s}$, but it does provide a unique relation between the initial 
entropy density profile and the final hadron multiplicity. The predicted 
increase of the final charged multiplicity from various models ranges from 
less than twice to more than 4 times the multiplicity at RHIC 
\cite{Kharzeev:2004if, Note1}. We therefore present our results as a 
function of final charged multiplicity, parameterizing it through the 
initial peak entropy density $s_0$ at thermalization time $\tau_0$ in 
$b{\,=\,}0$ collisions. We cover the range from 
$\frac{dN_{\mathrm{ch}}}{dy}(y{=}b{=}0){\,=\,}680$ 
(RHIC) to $\frac{dN_{\mathrm{ch}}}{dy}(y{=}b{=}0){\,=\,}2040$ (LHC). 

We use an EoS that transitions from an ideal quark-gluon plasma phase 
above a critical temperature $T_\mathrm{c}$ to a chemically 
non-equilibrated hadron resonance 
gas below $T_\mathrm{c}$ (RappEoS \cite{Kolb:2000sd}). The final hadron 
yields are assumed to freeze out directly at $T_\mathrm{c}{\,=\,}164$\,MeV.
The hadron momenta are taken to decouple at energy density 
$e_\mathrm{dec}{\,=\,}0.075$\,GeV/fm$^3$, $T_\mathrm{dec}{\,=\,}100$\,MeV.
Spectra and HBT radii are calculated from the emission function obtained 
via the Cooper-Frye prescription \cite{Cooper:1974mv} along this 
decoupling surface. For simplicity, we compute the HBT radii from
the space-time variances of this emission function instead of doing a 
Gaussian fit to the two-pion correlation function, knowing that this 
overestimates the longitudinal radius $R_l$ by 20-25\% 
\cite{Frodermann:2006sp}. A corresponding overall downward correction 
should thus be applied to all $R_l$ values shown below.

As the initial entropy density $s_0$ and temperature $T_0$ increase we 
reduce the therma\-li\-zation time $\tau_0$, keeping $T_0\tau_0$ constant. 
This yields a reduction from $\tau_0=0.6$\,fm/$c$ for 
$s_0{\,=\,}117$\,fm$^{-3}$, $\frac{dN_{\mathrm{ch}}}{dy}{\,=\,}680$ 
(``RHIC initial conditions'') 
to $\tau_0{\,=\,}0.35$\,fm/$c$ for $s_0{\,=\,}602$\,fm$^{-3}$, 
$\frac{dN_{\mathrm{ch}}}{dy}{\,=\,}2040$ (``LHC initial conditions''). 
%
%%%%%%%%%%%%%%%%%%%%%%%%%%%%%%% Fig. 1 %%%%%%%%%%%%%%%%%%%%%%%%%%%%%%%
\begin{figure}[t]
  \begin{center}
  \includegraphics[bb=7 28 789 590,width=\linewidth]{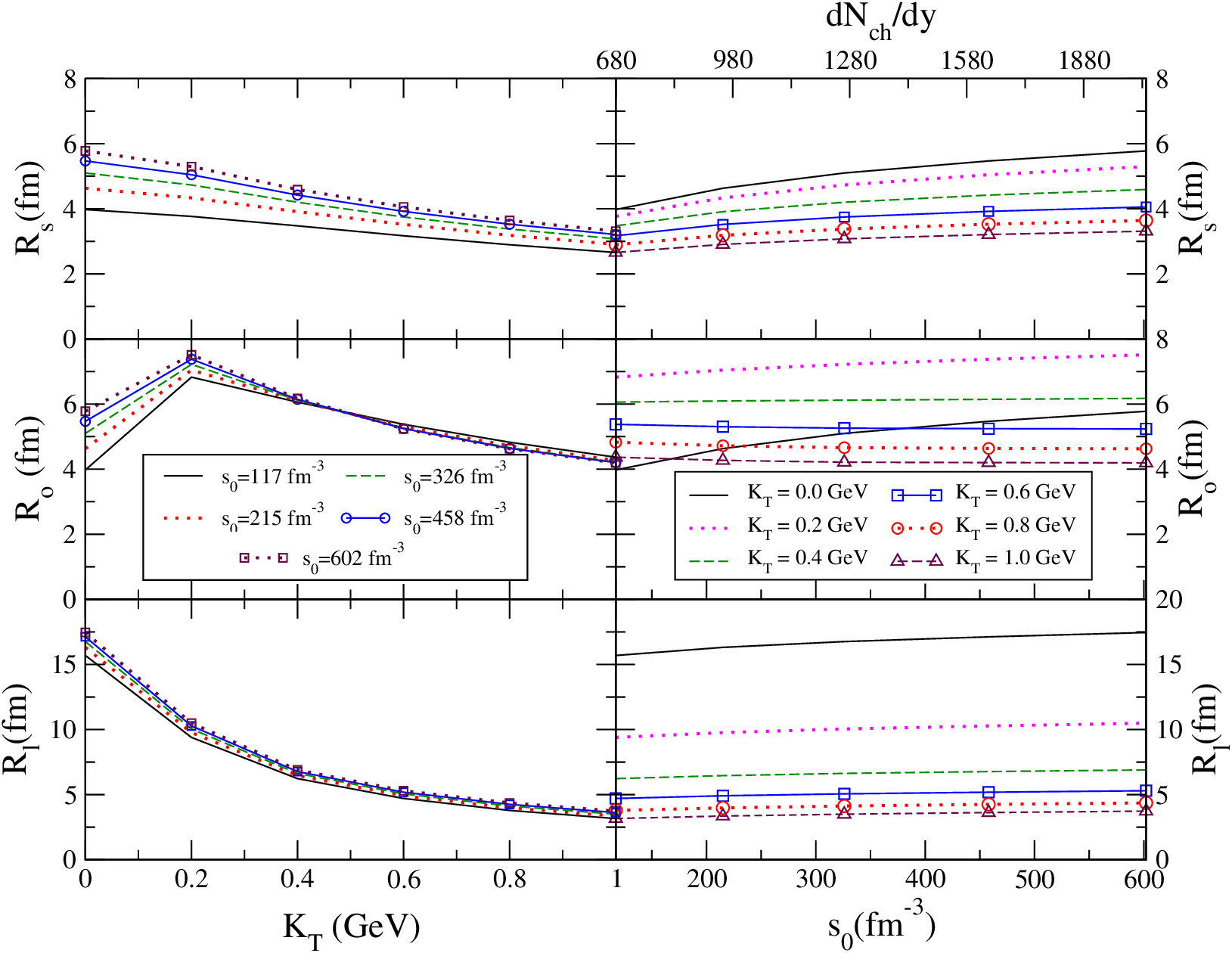}
  \end{center}
   \caption{\label{radiiKT}(Color online)
   Pion HBT radii for central ($b{=}0$) Au+Au collisions as a function of 
   transverse pair momentum $K_T$ (left) and of initial entropy density 
   $s_0$ or final charged multiplicity $\frac{dN_{\mathrm{ch}}}{dy}$ (right).} 
\end{figure}
%%%%%%%%%%%%%%%%%%%%%%%%%%%%%%%%%%%%%%%%%%%%%%%%%%%%%%%%%%%%%%%%%%%%%%
%

%%%%%%%%%%%%%%%%%%%%%%%%%%%%%%%%%%%%%%%%%%%%%%%%%%%%%%%%%%%%%%%%%%%%%%
%\section{Pion HBT radii}
%%%%%%%%%%%%%%%%%%%%%%%%%%%%%%%%%%%%%%%%%%%%%%%%%%%%%%%%%%%%%%%%%%%%%%
\section{Central collisions}
%%%%%%%%%%%%%%%%%%%%%%%%%%%%%%%%%%%%%%%%%%%%%%%%%%%%%%%%%%%%%%%%%%%%%%
\Fref{radiiKT} shows the pion HBT radii for central Au+Au (Pb+Pb) collisions 
in the out-side-long coordinate system \cite{concepts} as functions of 
transverse momentum and total charged multiplicity. There are no dramatic
changes, neither in magnitude nor in $K_T$-dependence, of the HBT radii as
we increase the charged multiplicity by up to a factor 3. The largest 
increase is seen for $R_s$ (by $\sim30\%$ at low $K_T$) while $R_o$
(which follows $R_s$ at $K_T{\,=\,}0$ by symmetry) even slightly 
decreases at large $K_T$. $R_l$ changes hardly at all. For comparison we 
also performed calculations with a transitionless ideal massless gas EoS 
\cite{Heinz:2002sq} (not shown graphically). In this case we see the smallest 
($<10\%$) increase in $R_s$, about 10-15\% increase in $R_o$ (at all $K_T$), 
and about 25-30\% increase in $R_l$. None of these changes will be easy to 
measure, but one sees that differences in the small predicted changes 
depend on the EoS, in particular on whether or not it embodies a phase 
transition. The main deficiency of hydrodynamic predictions for the HBT
radii at RHIC (too weak $K_T$-dependence of $R_s$ and $R_o$ and a ratio
$R_o/R_s$ much larger than 1) is not likely to be resolved at the 
LHC unless future LHC data completely break with the systematic tendencies 
observed so far \cite{Lisa:2005dd}.

%%%%%%%%%%%%%%%%%%%%%%%%% Fig. 2 %%%%%%%%%%%%%%%%%%%%%%%%%%%%%%%%%%%%%%%%%
\begin{figure}[tb]
  \begin{center}
  \includegraphics[width=\linewidth]{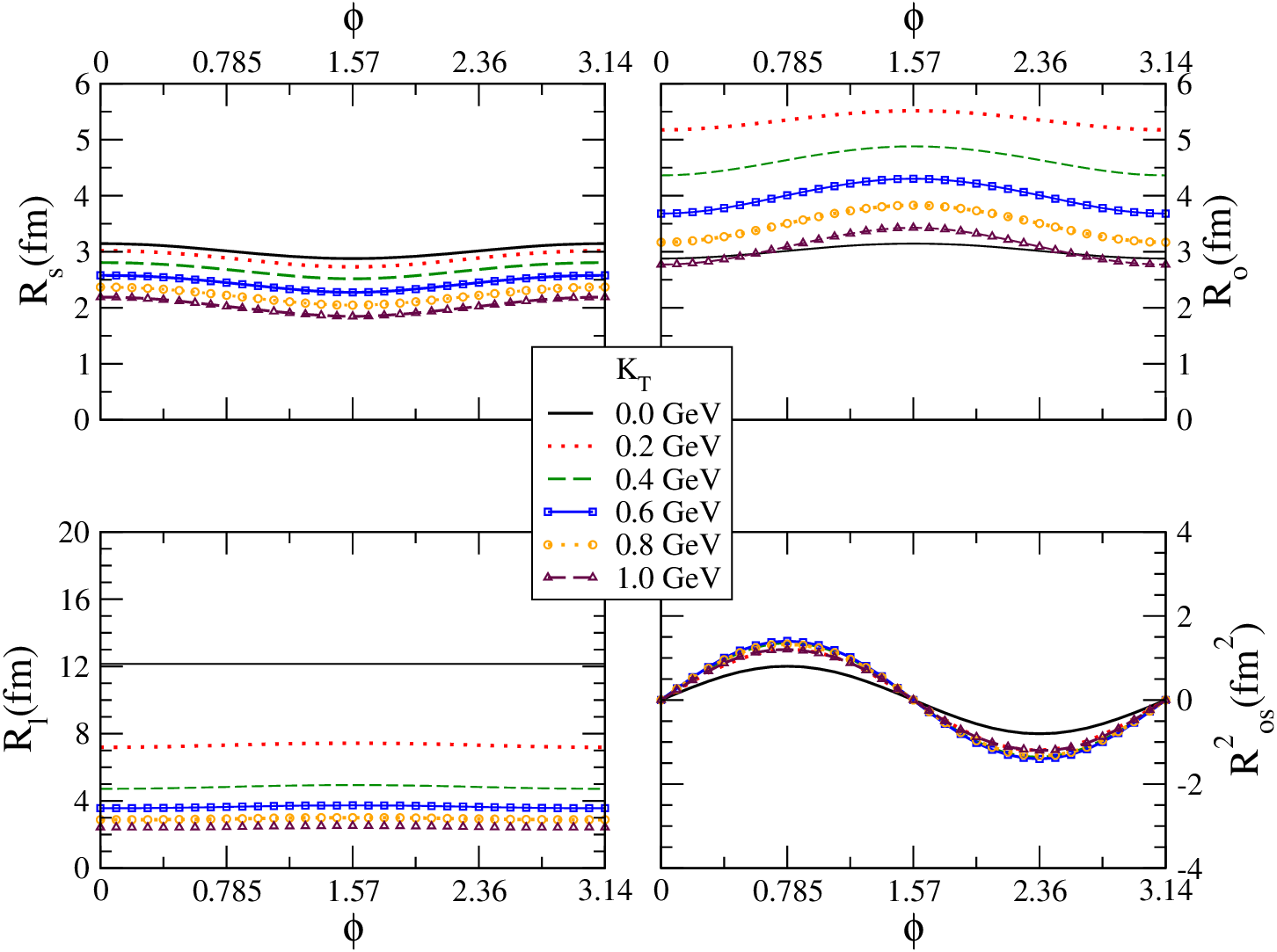}
  \end{center}
  \caption{\label{oscillationRHIC}(Color online)
  Pion HBT radii for non-central ($b{\,=\,}7$\,fm) Au+Au collisions as 
  a function of azimuthal angle $\Phi$ for different pair momenta $K_T$,
  for ``RHIC'' initial conditions (see text).
  } 
 \end{figure}
%%%%%%%%%%%%%%%%%%%%%%%%%%%%%%%%%%%%%%%%%%%%%%%%%%%%%%%%%%%%%%%%%%%%%%%%%%%

%%%%%%%%%%%%%%%%%%%%%%%%%%%%%%%%%%%%%%%%%%%%%%%%%%%%%%%%%%%%%%%%%%%%%%%
\section{Non-central collisions}
%%%%%%%%%%%%%%%%%%%%%%%%%%%%%%%%%%%%%%%%%%%%%%%%%%%%%%%%%%%%%%%%%%%%%%%

One of the strengths of AZHYDRO is simulating anisotropic, non-central 
collisions. For RHIC initial conditions, although the magnitudes of the 
HBT radii in central collisions were not predicted accurately, their 
normalized oscillation amplitudes at small $K_T$ (which measure the 
source eccentricity at freeze-out \cite{Retiere:2003kf}) were 
correctly reproduced \cite{concepts}. Their extrapolation to LHC initial 
conditions may therefore have predictive power.

\begin{figure}[t]
  \begin{center}
  \includegraphics[width=0.7\linewidth,clip=]{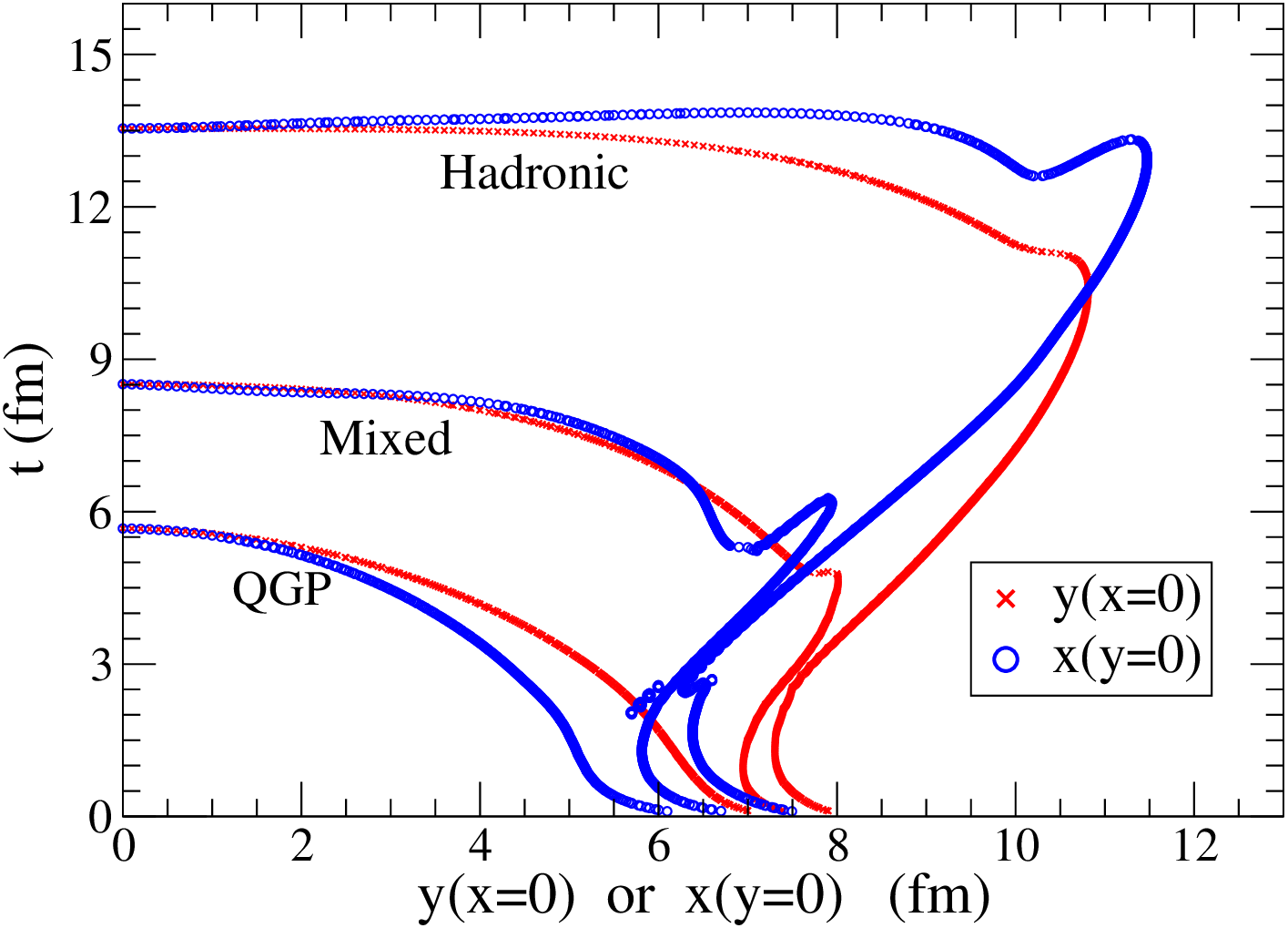}
  \end{center}
  \caption{\label{contours}(Color online)
  Cuts along the $x$ and $y$ axes through the freeze-out surface
  for (modified) ``LHC'' initial conditions: 
  $\frac{dN_\mathrm{ch}}{dy}{\,=\,}2035$, $s_0{\,=\,}1735$\,fm$^{-3}$
  at $\tau_0{\,=\,}0.1$\,fm/$c$ (this particular calculation was done
  for computation of photons and dileptons, using a reduced thermalization
  time to also capture some of the pre-equilibrium radiation). 
  }
\end{figure}

In Figures \ref{oscillationRHIC} and \ref{oscillationLHC} we show the 
azimuthal oscillations of the HBT radii for semiperipheral Au+Au
collisions at $b{\,=\,}7$\,fm, for both ``RHIC'' and ``LHC'' initial 
conditions defined above. At low pair momentum $K_T$, both $R_s$ and 
$R_o$ show an inversion of the sign of the oscillation amplitude between
RHIC and LHC. This is indicative of a transition from out-of-plane 
to in-plane deformation of the source at freeze-out \cite{Heinz:2002sq}.
\Fref{contours} shows cuts through the decoupling surface (as well as 
through the hypersurfaces indicating the transition from QGP to mixed 
phase and from mixed phase to hadron gas) along the $x$ and $y$ axes
(in-plane and out-of-plane, respectively) for ``LHC'' initial conditions: 
one observes that the source is initially wider in the $y$-direction 
(out-of-plane), but later becomes larger in the $x$-direction (in-plane). 
With an ideal gas EoS this shape transition requires much higher initial
entropy densities and temperatures \cite{Heinz:2002sq}.

%%%%%%%%%%%%%%%%%%%%%%%%% Fig. 3 %%%%%%%%%%%%%%%%%%%%%%%%%%%%%%%%%%%%%%%%%%
%%%%%%%%%%%%%%%%%%%%%%%%%%%%%%%%%%%%%%%%%%%%%%%%%%%%%%%%%%%%%%%%%%%%%%%%%%%
 
\begin{figure}
  \begin{center}
  \includegraphics[width=\linewidth]{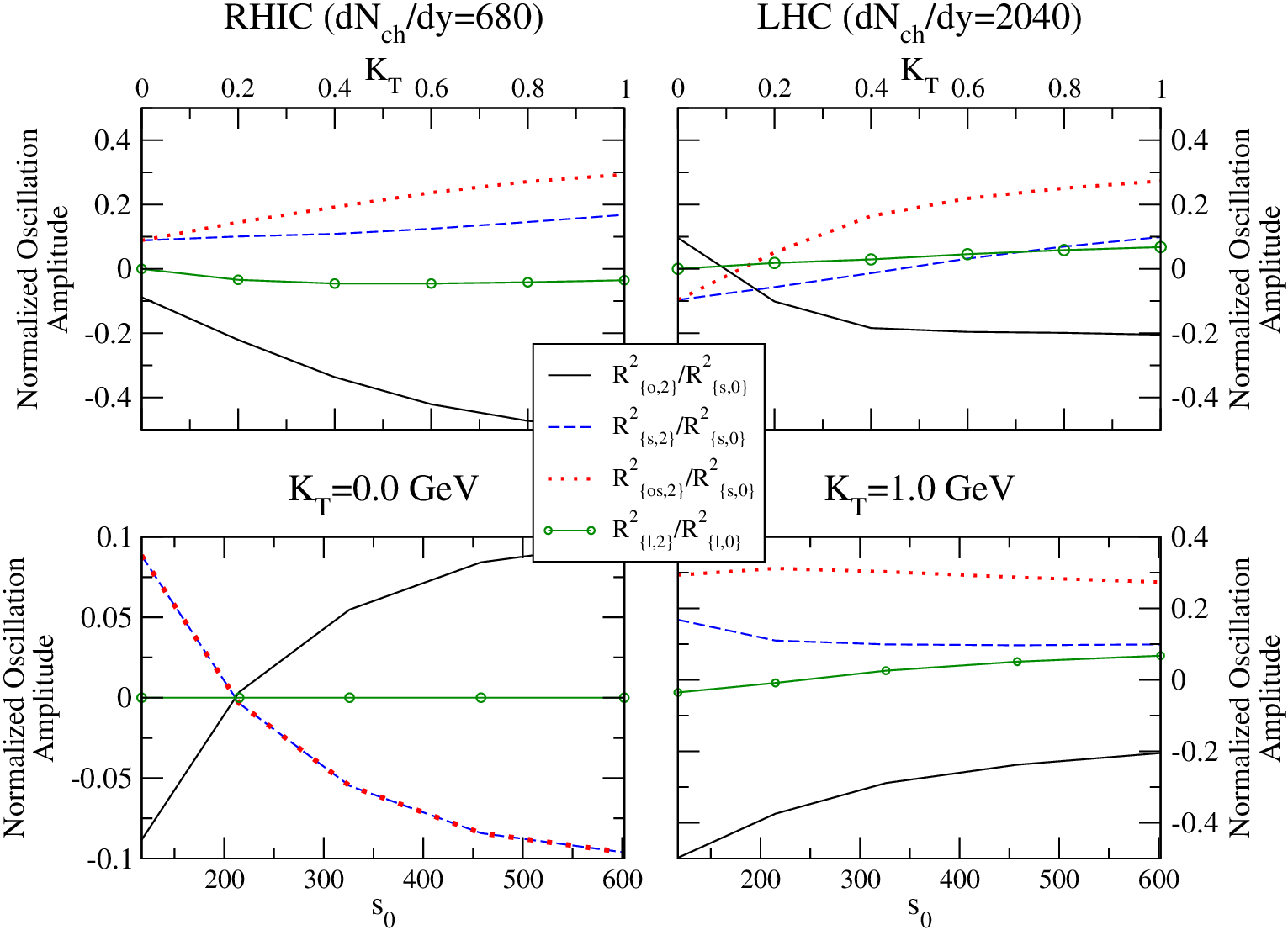}
  \end{center}
  \caption{\label{osciAmp}(Color online)
   Normalized HBT oscillation amplitudes as a function of $K_T$ at RHIC 
   and LHC (top) and as function of $s_0$ for two values of $K_T$ (bottom).
  }
\end{figure}

Retiere and Lisa \cite{Retiere:2003kf} introduced the normalized 
$2^{\mathrm{nd}}$ order Fourier components 
\begin{eqnarray}
  \frac{R^2_{(o,s),2}}{R^2_{s,0}} = 
  \frac{R^2_{(o,s)}(0) - R^2_{(o,s)}(\frac{\pi}{2})}
       {R^2_{s}(0)     + R^2_{s}(\frac{\pi}{2})}, 
\\
  \frac{R^2_{os,2}}{R^2_{s,0}} = 
  \frac{R^2_{(os)}(\frac{\pi}{4}) - R^2_{(os)}(\frac{3\pi}{4})}
       {R^2_{s}(0)                + R^2_{s}(\frac{\pi}{2})}, \qquad
  \frac{R^2_{l,2}}{R^2_{l,0}} = 
  \frac{R^2_{l}(0) - R^2_{l}(\frac{\pi}{2})}
       {R^2_{l}(0) + R^2_{l}(\frac{\pi}{2})}
\nonumber
\end{eqnarray}
of $R^2_{o,s,l}(\Phi)$, shown in \Fref{osciAmp} as functions of $K_T$
and $s_0$. They showed \cite{Retiere:2003kf} that the zero-momentum limit 
of $R^2_{s,2}/R^2_{s,0}$ is a direct measure of the spatial source 
eccentricity at freeze-out:
% 
%%%%%%%%%%%%%%%%%%%%%%%%% Fig. 4 %%%%%%%%%%%%%%%%%%%%%%%%%%%%%%%%%%%%%%%%%%
\begin{figure}[t]
  \begin{center}
  \includegraphics[width=\linewidth]{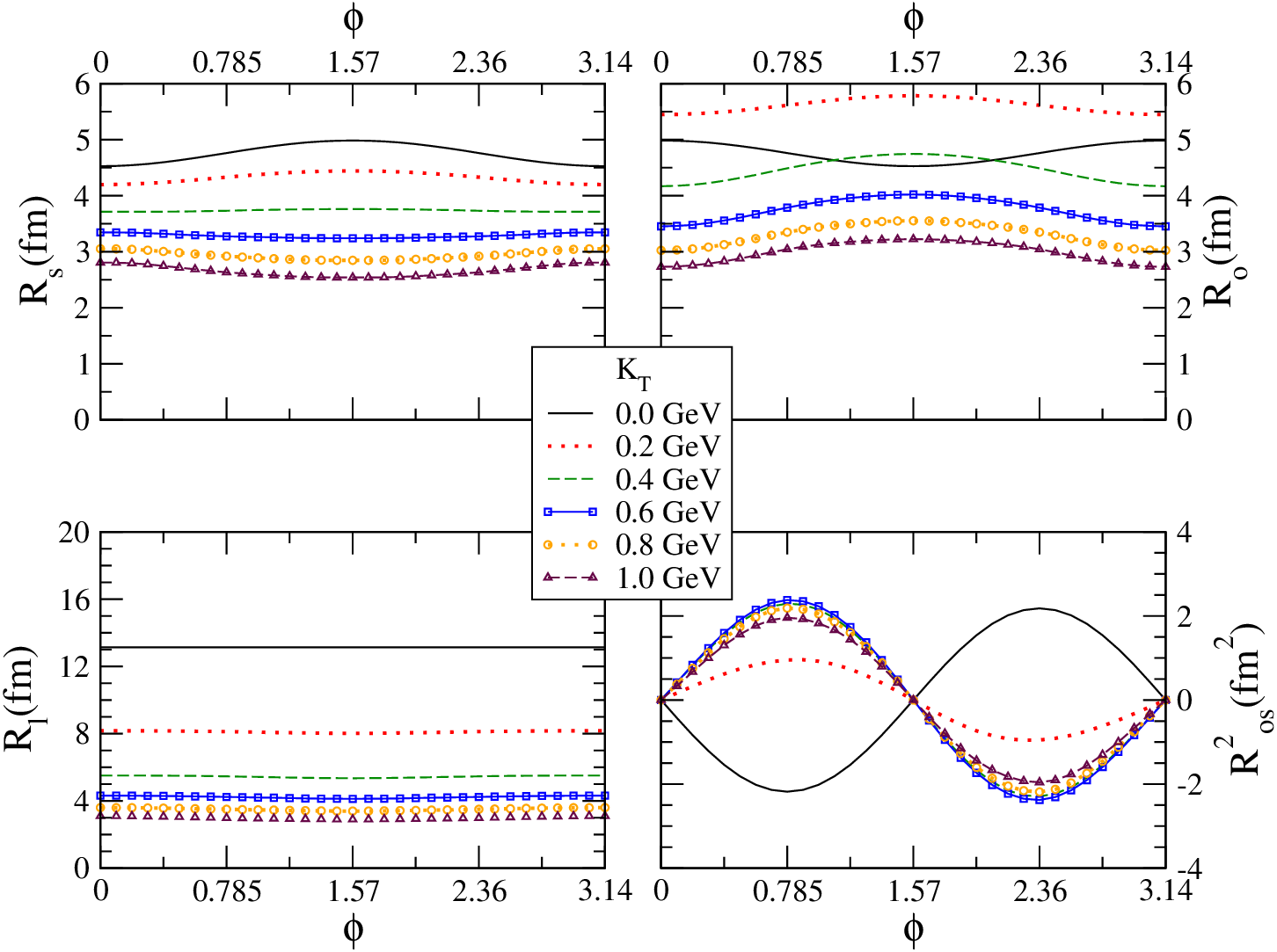}
  \end{center}
  \caption{\label{oscillationLHC}(Color online)
   Same as \Fref{oscillationRHIC}, but for ``LHC'' initial conditions.}
\end{figure}
%%%%%%%%%%%%%%%%%%%%%%%%%%%%%%%%%%%%%%%%%%%%%%%%%%%%%%%%%%%%%%%%%%%%%%%%%%%
%
$\epsilon_x^\mathrm{fo}{\,=\,}2\lim_{K_T\to0}\left(R^2_{s,2}/R^2_{s,0}\right)$.
Using this relationship, the lower left panel in \Fref{osciAmp} shows
that the freeze-out source eccentricity flips sign between RHIC and
LHC, and that at LHC the freeze-out source is elongated in the reaction
plane direction by about as large a factor as it was still out-of-plane
elongated in the RHIC case. This is different in runs with an ideal massless
gas EoS where even with ``LHC initial conditions'' the final freeze-out
source is found to be still out-of-plane elongated (although just barely so).

%%%%%%%%%%%%%%%%%%%%%%%%%% Fig.5 %%%%%%%%%%%%%%%%%%%%%%%%%%%%%%%%%%%%%%%%%% 
%%%%%%%%%%%%%%%%%%%%%%%%%%%%%%%%%%%%%%%%%%%%%%%%%%%%%%%%%%%%%%%%%%%%%%%%%%%

\section{Conclusions}
By varying the initial entropy density to control the final charged 
multiplicity, we used the hydrodynamic model to predict trends for the
pion HBT radii from Au+Au or Pb+Pb collisions as one moves from RHIC to 
LHC energies. In spite of a documented failure of the hydrodynamic model
to reproduce the HBT radii measured at RHIC, the model has had great success 
for most other soft-hadron observables, so the predicted trends may still be 
trustworthy. We find very little variation in the HBT radii for central
collisions, and whatever small differences we see depends sensitively
on details of the equation of state, in particular whether or not it 
embodies a quark-hadron phase transition. Clear and characteristic changes
are predicted for the normalized azimuthal oscillation amplitudes of
the HBT radii from non-central collisions, indicative of a qualitative
change of the shape of the source at freeze-out which evolves from 
an out-of-plane elongated freeze-out configuration at RHIC to an in-plane
elongated shape at the LHC.\\[1ex]

\noindent
{\bf Acknowledgements:} This work was supported by the U.S. Department
of Energy under grant DE-FG02-01ER41190 and a University Presidential 
Fellowship from The Ohio State University (E.F.).

%%%%%%%%%%%%%%%%%%%%%%%%%%%%%%%%%%%%%%%%%%%%%%%%%%%%%%%%%%%%%%%%%%%%%%%%%%

\end{document}